\documentclass[conference]{IEEEtran}
\usepackage{enumitem}
\usepackage{amsmath,amssymb,amsthm}
\usepackage{centernot}
\usepackage{graphicx} 
\usepackage{hyperref}
\usepackage{overpic}

\newcommand{\defeq}{\triangleq}
\newcommand{\hmin}{H_{\rm{min} } }
\newcommand{\pguess}{p_{\operatorname{guess} } }
\newcommand{\qcorr}{q_{\operatorname{corr}}}
\newcommand{\set}[1]{\mathsf{#1}}

\newcommand{\hil}[1]{\mathcal{#1}}
\newcommand{\id}{\operatorname{id}}

\newcommand{\Tr}[1]{{\operatorname{Tr}}\!\left[{#1}\right]}
\newcommand{\PTr}[2]{\operatorname{Tr}_{#1}\!\left[{#2}\right]}

\renewcommand{\>}{\rangle}
\newcommand{\<}{\langle}
\newcommand{\A}{{\bar{A}}}
\newcommand{\mE}{\mathcal{E}}
\newcommand{\R}{\bar{R}}
\newcommand{\mN}{\mathcal{N}}
\newcommand{\mD}{\mathcal{D}}

\renewcommand{\vec}[1]{\pmb{\mathrm{#1}}}

\newcommand{\sH}{\hil{H}}

\newtheorem{theorem}{Theorem}
\newtheorem{corollary}{Corollary}

\newtheorem{definition}{Definition}

\theoremstyle{remark}

\ifCLASSINFOpdf
\else
\fi
\begin{document}
%
% paper title
% Titles are generally capitalized except for words such as a, an, and, as,
% at, but, by, for, in, nor, of, on, or, the, to and up, which are usually
% not capitalized unless they are the first or last word of the title.
% Linebreaks \\ can be used within to get better formatting as desired.
% Do not put math or special symbols in the title.
\title{Comparison of Noisy Channels\\ and Reverse Data-Processing Theorems}

% author names and affiliations
% use a multiple column layout for up to three different
% affiliations
\author{\IEEEauthorblockN{Francesco Buscemi}
\IEEEauthorblockA{Graduate School of Informatics, Nagoya University\\
Furo-cho, Chikusa-ku, Nagoya, 464-8601 Japan\\
Email: buscemi@is.nagoya-u.ac.jp}}

\maketitle

\begin{abstract}
This paper considers the comparison of noisy channels from the viewpoint of statistical decision theory. Various orderings are discussed, all formalizing the idea that one channel is ``better'' than another for information transmission. The main result is an equivalence relation that is proved for classical channels, quantum channels with classical encoding, and quantum channels with quantum encoding.
\end{abstract}
\bigskip

\IEEEpeerreviewmaketitle

\begin{figure}[t]
	\centering
		\begin{overpic}[width=\linewidth]{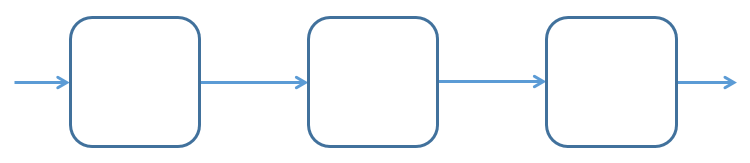}
		\put (4, 12) {\large $u$}
		\put (12, 9) {{\large $p(x|u)$}}
		\put (32, 12) {\large $x$}
		\put (42.5, 9) {{\large $w_1(y|x)$}}
		\put (64, 12) {\large $y$}
		\put (74.5, 9) {{\large $w_2(z|y)$}}
		\put (93, 12) {\large $z$}
	\end{overpic}
	\caption{Shannon's basic communication scheme: a message labeled by $u$ is passed through the (probabilistic) encoding $p(x|u)$, which produces the input $x$ to the channel. This is transmitted to the receiver via the communication channel $w_1$. The receiver obtains the output symbol $y$ and decodes it with $w_2$, thus obtaining an estimate $z$ of $u$.}
	\label{fig:shannon-scheme}
\end{figure}
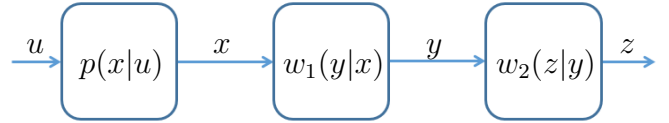

A \textit{data-processing inequality} is a mathematical statement formalizing the fact that the information content of a signal cannot be increased by post-processing. One of the simplest scenarios in which a data-processing inequality can be formulated is the following~\cite{cziszar-korner,cover-thomas}. Given are two noisy channels $w_1:\set{X}\to\set{Y}$ and $w_2:\set{Y}\to\set{Z}$, where $\set{X}=\{x\}$, $\set{Y}=\{y\}$ , and $\set{Z}=\{z\}$ are three finite alphabets. Then, for any index set $\set{U}=\{u\}$ and any initial joint distribution $p(x,u)$, the joint distribution $\sum_xw_2(z|y)w_1(y|x)p(x,u)$ satisfies the following inequalities:
\begin{equation}\label{eq:data-proc}
I(U;Y)\ge I(U;Z)\;.
\end{equation}
Interpreting the random variable $U$ as the (index labeling the) message, $X$ as the transmitted signal (codeword), $w_1$ as the communication channel, $Y$ as the output signal, $w_2$ as the decoding, and $Z$ as the recovered message, the above inequality formalizes the fact that the information content carried by the signal about the message cannot be increased by any decoding performed locally at the receiver (see Fig.~\ref{fig:shannon-scheme}).

Data-processing inequalities hence provide necessary conditions for the communication process to be ``local.'' Namely, data-processing inequalities must be obeyed whenever the physical process carrying the message from the sender to the receiver is composed by computationally isolated parts (encoding, transmission, decoding, etc.). Any information that is communicated must be transmitted via a physical signal: as such, in the absence of an external memory, information can only decrease, never increase, along the transmission. Hence, ``locality'' in this sense can be understood as the condition that the process $U\to X\to Y\to Z$ forms a Markov chain.
%in order to avoid confusion with other connotations of the word\footnote{In this work, memoryless process, Markov local process, and computationally isolated process are all synonyms. We prefer however to maintain all three terms because they in fact highlight different aspects of the same information-theoretic concept.}.

Here we aim to derive statements that provide \textit{sufficient} conditions for Markov locality, in the form of a set of information-theoretic inequalities. We refer to such statements as \textit{reverse data-processing theorems}. For example, suppose that, given two noisy channels $w:\set{X}\to\set{Y}$ and $w':\set{X}\to\set{Z}$, for any set $\set{U}$ and for any initial joint distribution $p(x,u)$, the resulting distributions $\sum_xw(y|x)p(x,u)$ and $\sum_xw'(z|x)p(x,u)$ always satisfy the inequality $I(U;Y)\ge I(U;Z)$. Can we then conclude that there exists a noisy channel $\varphi:\set{Y}\to\set{Z}$ such that $w'(z|x)=\sum_y\varphi(z|y)w(y|x)$?

\bigskip\section{Comparison of noisy channels}

An answer in the affirmative to the above question would constitute an example of a reverse data-processing theorem (RDPT). Clearly, a RDPT is a statement about the comparison of two noisy channels, formalizing the concept that one channel is ``more informative'' than the other one. This problem, first considered by Shannon~\cite{shannon_1958}, is intimately related to the theory of statistical comparisons~\cite{blackwell_equivalent_1953,torgersen_comparison_1991,cohen_comparisons_1998,liese-miescke}, even though, quite surprisingly, this connection remained unexplored until recently~\cite{Raginsky2011}.

As a matter of fact, the RDPT tentatively formulated above does not hold: as it has been shown by K\"orner and Marton in~\cite{Korner1977}, the ``noisiness ordering'' is only necessary, but not sufficient, for the ``degradability ordering.'' Such pre-orderings  (that become partial orderings when defined on channels equivalence classes) are defined as follows (here we follow~\cite{ElGamal1977}):
\begin{definition}[\cite{Korner1977,ElGamal1977}]\label{def:less-noisy}
	Given are two noisy channels, $w:\set{X}\to\set{Y}$ and $w':\set{X}\to\set{Z}$.
	\begin{enumerate}[label=\roman*)]
		\item The channel $w$ is said to be \emph{degradable} to $w'$ if and only if there exists another channel $\varphi:\set{Y}\to\set{Z}$ such that
		\begin{equation}\label{eq:degradable}
		w'(z|x)=\sum_y\varphi(z|y)w(y|x)\;.
		\end{equation}
		\item The channel $w$ is said to be \emph{less noisy} than $w'$ if and only if, for any index set $\set{U}$, any distribution $p(u)$, and any encoding $p(x|u)$, the resulting distributions $\sum_x w(y|x)p(x,u)$ and $\sum_x w'(z|x)p(x,u)$ always satisfy the inequality
		\begin{equation}\label{eq:less-noisy}
		H(U|Y)\le H(U|Z)\;.
		\end{equation}
%		\item The channel $w$ is said to be \emph{more capable} than $w'$ if and only if, for any distribution $p(x)$, the resulting distributions $w(y|x)p(x)$ and $w'(z|x)p(x)$ always satisfy the inequality
%		\begin{equation}\label{eq:more-cap}
%		H(X|Y)\le H(X|Z)\;.
%		\end{equation}
	\end{enumerate}
\end{definition}
It is immediate to see that the data-processing inequality~(\ref{eq:data-proc}) is equivalent to the relation (\textit{i})$\implies$(\textit{ii}), while the fact that a RDPT does not hold in this case is equivalent to the relation (\textit{ii})$\centernot{\implies}$(\textit{i}), proved in~\cite{Korner1977} by means of explicit counterexamples.

As there exist many data-processing inequalities (essentially, any meaningful measure of information should satisfy one), even though the RDPT does not hold for~(\ref{eq:data-proc}), it may still be possible that other data-processing inequalities instead satisfy the reverse property. Here we focus in particular on an alternative figure of merit for information-theoretic protocols, namely, the average error probability. We can hence reformulate the definition of noisiness ordering in terms of the average error probability, instead of the conditional entropy. We thus introduce an ``ambiguity ordering'' as follows:
\begin{definition}[Ambiguity]\label{def:less-mistake}
	The channel $w$ is said to be \emph{less ambiguous} than $w'$ if and only if, for any index set $\set{U}$, any distribution $p(u)$, and any encoding $p(x|u)$, the average error probabilities always satisfy the inequality
	\begin{equation}\label{eq:less-mistake}
	\pguess(U|Y)\ge\pguess(U|Z)\;,
	\end{equation}
	where $\pguess(U|Y)\defeq \sup_{d(u|y)}\sum_{x,y}d(u|y)w(y|x)p(x|u)p(u)$ and analogously for $\pguess(U|Z)\;$.
\end{definition}

The formal relation between Definitions~\ref{def:less-noisy} and~\ref{def:less-mistake} can be further clarified by noticing that, in terms of the conditional min-entropy $\hmin(U|Y)=-\log_2\pguess(U|Y)$~\cite{Ren05,KRS09},
\begin{align}\label{eq:perr-hmin-class}
&\pguess(U|Y)\ge\pguess(U|Z)\\
&\iff \hmin(U|Y)\le\hmin(U|Z)\;.\nonumber
\end{align}
In this form, the close analogy between noisiness and ambiguity is apparent. In what follows, we will focus on the ambiguity ordering and prove that, contrarily to the noisiness ordering, it is \textit{equivalent} to the degradability ordering. We will consider from the beginning the case of quantum channels, recovering the classical scenario in the completely commutative case.

\bigskip\section{Quantum channels}

In the quantum case, alphabets and distributions are replaced by (finite dimensional) Hilbert spaces and density operators, respectively. Hilbert spaces are denoted by $\sH_A,\sH_B,\sH_C,\dots$, density operators by $\rho,\sigma,\tau,\dots$. Formally, a density operator satisfies two properties: $\rho\ge0$ and $\Tr{\rho}=1$. When the system is in a pure state $|\varphi\>$, we denote the corresponding density operator by $\varphi\defeq|\varphi\>\<\varphi|$. Given two copies of the same system $A$, we denote the corresponding Hilbert spaces by $\sH_A$ and $\sH_{\bar{A}}\cong\sH_A$. We also need to fix one privileged maximally entangled state that we denote by $|\Phi^+_{\A A}\>$, and that in turns fixes two orthonormal bases once it is written as $|\Phi^+_{\A A}\>=d_A^{-1/2}\sum_i|i_\A\>|i_A\>\;$.

\begin{figure}
	\centering 
	\begin{overpic}[width=\linewidth]{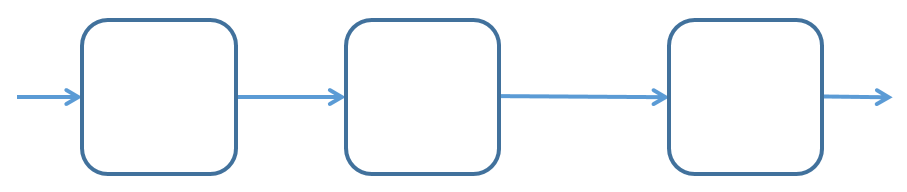}
		\put (4, 12) {\large $u$}
		\put (15, 8) {{\huge $\mE$}}
		\put (29.5, 12.5) {\large $\tau^u_A$}
		\put (42.5, 8) {{\huge $\mN$}}
		\put (58, 12.5) {\large $\mN(\tau^u_A)$}
		\put (79, 8) {{\huge $\mathcal{D}$}}
		\put (93, 12) {\large $\hat{u}$}
	\end{overpic}
	\caption{The basic communication scenario in which classical information is conveyed through a quantum channel. A classical message, indexed by $u$, in encoded on the quantum state $\tau^u_A$, which is then fed through the channel $\mN$. The receiver, upon receiving the output state $\mN(\tau^u_A)$ performs a measurement (i.e., applies a quantum-classical decoding) in order to guess the value of $u$.}
	\label{fig:figure-cq}
\end{figure}

\begin{figure}
	\centering
	\begin{overpic}[width=\linewidth]{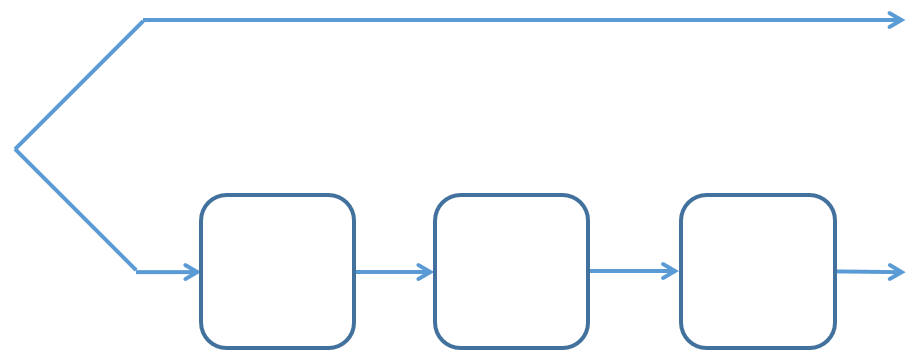}
		\put (16, 11) {$R$}
		\put (27.5, 7) {{\huge $\Gamma$}}
		\put (41, 11) {$A$}
		\put (52, 7) {{\huge $\mN$}}
		\put (68, 11) {$B$}
		\put (80, 7) {{\huge $\mathcal{D}$}}
		\put (94, 11) {$R$}
		\put (94, 32) {$\bar{R}$}
		\put (5, 22) {{\large $\varphi_{R\bar{R}}$}}
	\end{overpic}
	\caption{A typical fully quantum communication scenario. The ``message'' now is the entanglement in the initial state $|\varphi_{\R R}\>$. The $R$ subsystem is passed through an encoding channel $\Gamma:R\to A$, and then transmitted to the receiver through $\mN:A\to B$. The receiver then locally applies a decoding channel $\hil{D}:B\to R$, trying to maximize the overlap of the overall bipartite state with the maximally entangled state $|\Phi^+_{\R R}\>$.}
	\label{fig:figure-qq}
\end{figure}

A quantum communication channel is described by a completely positive trace-preserving (CPTP) linear map $\mN$, mapping linear operators on the input space $\sH_A$ to linear operators on the output space $\sH_B$. In this case we will simply write $\mN:A\to B$. The identity channel will be denoted by $\id$. Since a quantum channel can be used to convey classical or quantum information, we have two natural ways to generalize Def.~\ref{def:less-mistake}, schematically represented in Figs.~\ref{fig:figure-cq} and~\ref{fig:figure-qq}.

At this point we need to recall the definition of quantum conditional min-entropy~\cite{Ren05}:
\begin{equation*}
\hmin(A|B)_\rho \defeq \log_2\inf_{\sigma_B \ge 0} \{ \Tr{\sigma_B} : \rho_{AB}\le I_A\otimes\sigma_B\}\;.
\end{equation*}
The quantum conditional min-entropy is known to have many applications in one-shot quantum information theory, recovering the usual von Neumann conditional entropy in a suitable limit (see, e.g., Ref.~\cite{Toma-book}). In particular, in the case in which the bipartite state $\rho_{AB}$ is classical-quantum, i.e., $\rho_{AB}=\sum_up(u)|u\>\<u|_U\otimes\rho^u_B$, the corresponding conditional min-entropy $\hmin(U|B)_\rho$ is related to the error probability of guessing $U$ given $B$ as follows~\cite{KRS09}:
\begin{align*}
\hmin(U|B)_\rho&=-\log_2\sup_{\{P^u_B\}:\text{ POVM}}\sum_up(u)\Tr{\rho^u_B\ P^u_B}\\
&\defeq-\log_2 \pguess(U|B)\;,
\end{align*}
in perfect analogy with Eq.~(\ref{eq:perr-hmin-class}). For a general bipartite state $\rho_{AB}$ we have the following:
\begin{align*}
&\hmin(A|B)_\rho\\
&=-\log_2\sup_{\mD:\text{ CPTP}}d_A\<\Phi_{A \bar{A}}^+|(\id_{A}\otimes\mD_B)(\rho_{AB})|\Phi_{A\bar{A}}^+\>\\
&\defeq -\log_2\qcorr(A|B)\;.
\end{align*}

We are now ready to state the definition. (We refer the reader to Ref.~\cite{Wat12} for a different ordering.)
\begin{definition}[Ambiguity and Coherence]\label{def:q-less-noisy}
	Given are two quantum channels, $\mN:A\to B$ and $\mN':A\to B'$.
	\begin{enumerate}[label=\roman*)]
		\item The channel $\mN$ is said to be \emph{degradable} to $\mN'$ if and only if there exists another quantum channel $\Psi:B\to B'$ such that
		\begin{equation}\label{eq:q-degradable}
		\mN'=\Psi\circ\mN.
		\end{equation}
		\item The channel $\mN$ is said to be \emph{less ambiguous} than $\mN'$ if and only if, for any set $\set{U}$, any distribution $p(u)$, and any cq-encoding channel $\hil{E}:\set{U}\to A$, $u\mapsto\tau^u_A\;$, the resulting joint states $\rho_{UB}\defeq\sum_up(u)|u\>\<u|_U\otimes\mN(\tau^u_A)$ and $\sigma_{UB'}\defeq\sum_up(u)|u\>\<u|_U\otimes\mN'(\tau^u_A)$ always satisfy the inequality $\pguess(U|B)\ge\pguess(U|B')$, that is,
		\begin{equation}\label{eq:q-less-noisy}
		\hmin(U|B)_\rho\le \hmin(U|B')_\sigma\;.
		\end{equation}
		\item The channel $\mN$ is said to be \emph{more coherent} than $\mN'$ if and only if, for any auxiliary quantum system $R$, any bipartite pure state $|\varphi_{\R R}\>$ on $\sH_{\R}\otimes\sH_R\cong\sH_R^{\otimes 2}$, and any quantum encoding channel $\Gamma:R\to A$, the resulting bipartite states $\rho_{\R B}\defeq(\id_{\R}\otimes\mN_A\circ\Gamma_R)\varphi_{\R R}$ and $\sigma_{\R B'}\defeq(\id_{\R}\otimes\mN'_A\circ\Gamma_R)\varphi_{\R R}$ always satisfy the inequality $\qcorr(\bar{R}|B)\ge\qcorr(\bar{R}|B')$, that is,
		\begin{equation}\label{eq:q-comp-less-noisy}
		\hmin(\R|B)_\rho\le \hmin(\R|B')_\sigma\;.
		\end{equation}
	\end{enumerate}
\end{definition}

\bigskip\section{Main result: equivalence relations}

We are now ready to state the main result.

\begin{theorem}\label{th:main}
	Given are two quantum channels $\mN:A\to B$ and $\mN':A\to B'$. The following are equivalent.
	\begin{enumerate}[label=\roman*)]
		\item The channel $\mN$ is degradable to $\mN'$.
		\item For any auxiliary quantum system $C$, the extension $\id_C\otimes\mN_A$ is less ambiguous than the extension $\id_C\otimes\mN'_A\;$.
		\item For some auxiliary quantum system $C\cong B'$, and for some bijective\footnote{A \textit{bijective} channel here is meant as a CPTP map that preserves linear independence.} channel $\mathcal{M}:C\to C$, the extension $\hil{M}_C\otimes\mN_A$ is less ambiguous than $\hil{M}_C\otimes\mN'_A$.
		\item The channel $\mN$ is more coherent than $\mN'$.
		\item For some auxiliary quantum system $R\cong\R\cong B'$, the bipartite states $\rho_{\R B}=(\id_{\R}\otimes\mN_A\circ\Gamma_R)\Phi^+_{\R R}$ and $\sigma_{\R B'}=(\id_{\R}\otimes\mN'_A\circ\Gamma_R)\Phi^+_{\R R}$ satisfy the inequality
		\[
		\hmin(\R|B)_\rho\le \hmin(\R|B')_\sigma\;,
		\]
		for any entanglement breaking quantum channel $\Gamma:R\to A\;$.
	\end{enumerate}
\end{theorem}

Before proceeding with a sketch proof, a few comments are in order. Condition~(\textit{iii}) essentially states that condition~(\textit{ii}) is redundant: it is not necessary to check that $\id_C\otimes\mN_A$ is less ambiguous than $\id_C\otimes\mN'_A$ \textit{for all} extensions $C$, but it is enough to consider just one particular extension. Moreover, there is no need to consider the identity channel on $C$, but any bijective channel (even an entanglement-breaking one) will do.

In the same way, condition~(\textit{v}) states that~(\textit{iv}) is redundant, in that it is not necessary to check the maximally entangled fraction \textit{for all} initial pure bipartite states $|\varphi_{\R R}\>$, but looking only at what the composed channel $\mN\circ\Gamma$ does on the maximally entangled reference state $|\Phi^+_{\R R}\>$ is enough. Moreover, we do not need to consider all possible encodings $\Gamma:R\to A$, but to check the condition only for entanglement-breaking encodings is sufficient. This is somewhat surprising given that the coherence ordering is defined in terms of the maximally entangled fraction.

The problem of characterizing statistical relations equivalent to degradability (and approximate degradability) has been considered before~\cite{clean-povm,shmaya,chefles,buscemi-qblackwell,BDS15,jencova-rand-channel,buscemi-divisibility,buscemi-prob-inf-trans, jencova-approx}, but only in the classical-quantum scenario (i.e., that of Fig.~\ref{fig:figure-cq}). A fully quantum scenario, alternative to the one proposed here, is discussed in~\cite{future}.

In the special case in which the channel $\mN'$ has a classical output, in the sense that for any input the output is always diagonalizable in some fixed basis, Theorem~\ref{th:main} can be strengthened as follows:
\begin{theorem}[\cite{buscemi-qblackwell,buscemi-prob-inf-trans}]\label{th:commuting}
	If the output of channel $\mN'$ is commuting, then $\mN$ is degradable to $\mN'$ if and only if $\mN$ is less ambiguous than $\mN'$. In other words, in this case it is not necessary to consider extensions of the channels.
\end{theorem} 

Moreover, point~(\textit{v}) of Theorem~\ref{th:main} implies the following result for classical channels:

\begin{corollary}[Classical channels]
	Given two noisy channels $w:\set{X}\to\set{Y}$ and $w':\set{X}\to\set{Z}$, the following are equivalent.
	\begin{enumerate}[label=\roman*)]
		\item The channel $w$ is degradable to $w'$.
		\item The channel $w$ is less ambiguous than $w'$.
		\item For any encoding $p(x|u)$, the joint distributions $\frac{1}{|\set{U}|}\sum_xw(y|x)p(x|u)$ and $\frac{1}{|\set{U}|}\sum_xw'(z|x)p(x|u)$ always satisfy $\pguess(U|Y)\ge \pguess(U|Z)\;$. In fact, it is possible to fix $\set{U}=\set{Z}$, without loss of generality.
	\end{enumerate}
\end{corollary}

In other words, in Def.~\ref{def:less-mistake} it is possible to take, without loss of generality, $\set{U}\equiv\set{Z}$ and uniform $p(u)$, and still have equivalence between ambiguity ordering and degradability ordering.

\section{Sketch proof of theorem~\ref{th:main}}

The implications (\textit{i})$\implies$(\textit{ii})$\implies$(\textit{iii}) are easy to see. The implication (\textit{iii})$\implies$(\textit{i}) has been proved, using various techniques and in various degrees of generality, in Refs.~\cite{chefles,BDS15,jencova-rand-channel,buscemi-prob-inf-trans,jencova-approx}.

In the same way, the implications (\textit{i})$\implies$(\textit{iv})$\implies$(\textit{v}) are also easy to see. Hence, in order to close the logical loop, we only need to prove that (\textit{v})$\implies$(\textit{i}).

	Given two quantum channels $\mN:A\to B$ and $\mN':A\to B'$, let us consider the corresponding Choi operators, namely,
	\begin{align*}
	&\rho_{\bar{A}B}=(\id_{{\bar{A}}}\otimes\mN_A)\Phi^+_{\bar{A}A}\;,\\
	&\sigma_{\bar{A}B'}=(\id_{\bar{A}}\otimes\mN'_A)\Phi^+_{\bar{A}A}\;,
	\end{align*}
	where $\bar{A}$ is a system isomorphic to $A$ (i.e., $d_A=d_{\bar{A}}$) and $\Phi^+_{\bar{A}A}=|\Phi^+_{\bar{A}A}\>\<\Phi^+_{\bar{A}A}|$ is the projector on the maximally entangled state $d_A^{-1/2}\sum_i|i_{\bar{A}}\>|i_A\>$. (The bases $|i_{\bar{A}}\>$ and $|i_A\>$ are fixed here once and for all in this proof.) In terms of the Choi operators, statement~(\textit{i}) can be reformulated as the existence of a channel $\Psi:B\to B'$ such that
	\begin{equation}\label{eq:goal}
	\sigma_{\bar{A}B'}=(\id_{{\bar{A}}}\otimes\Psi_B)\rho_{\bar{A}B}\;.
	\end{equation}
	
	Let us now introduce a set of density operators $\{\omega^i_{\bar{A}} \}_i$ spanning the whole set of linear operator on $\sH_{\bar{A}}$. Then, Eq.~(\ref{eq:goal}) is equivalent to the following:
	\begin{equation}
	\sigma^i_{B'}=\Psi(\rho^i_B)\;,\quad\forall i\;,
	\end{equation}
	where
	\begin{align*}
	&\rho^i_B=d_\A \PTr{\bar{A}}{(\omega^i_{\bar{A}}\otimes I_B)\ \rho_{\A B}}\;,\\
	&\sigma^i_{B'}=d_\A \PTr{\bar{A}}{(\omega^i_{\bar{A}}\otimes I_{B'})\ \sigma_{\A B'}}\;.
	\end{align*}
	Notice that the operators $\rho^i_B$ and $\sigma^i_{B'}$ are all well-defined density operators, as a consequence of the fact that $\rho_{\A}=\sigma_{\A}=d_{\A}^{-1}I_{\A}$).
	
	Let us introduce now another spanning set $\{X^j_{B'} \}_j$ for the set of linear operators on $\sH_{B'}$. This time, however, the operators $X^j_{B'}$ need only be self-adjoint. Then, statement~(\textit{i}) becomes equivalent to the existence of a CPTP map $\Psi$ such that
	\begin{align*}
	\vec{s} =\vec{r}(\Psi)\;,
	\end{align*}
	where we defined the two real vectors $\vec{s}=(s_{ij})$ and $\vec{r}(\Psi)=(r_{ij}(\Psi))$ as follows:
	\begin{align*}
	&s_{ij}=\Tr{\sigma^i_{B'}\;X^j_{B'}}\;,\\
	&r_{ij}(\Psi)=\Tr{\Psi(\rho^i_B)\;X^j_{B'}}\;.
	\end{align*}
	By defining the set $\set{S}(\rho)=\{\vec{r}(\Psi)\;|\;\Psi:B\to B'\text{ CPTP} \}$, statement~(\textit{i}) in this notation can be equivalently expressed as
	\[
	\vec{s}\in\set{S}(\rho)\;.
	\]
	
	The crucial observation now is that the set $\set{S}(\rho)$ is closed, bounded, and convex, as it inherits this structure from the set of CPTP maps $\Psi$. Hence, as an application of the separation theorem for convex sets~\cite{Roc70}, we have that statement~(\textit{i}) is equivalent to the following: for any real vector $\vec{c}=(c_{ij})$,
	\[
	\vec{s}\cdot\vec{c}\le\max_{\vec{r}\in\set{S}(\rho)}\vec{r}\cdot\vec{c}\;.
	\]
	
	The next step is to define, for each choice of $\vec{c}$, the corresponding self-adjoint operators $Y^i_{B'}=\sum_jc_{ij}X^j_{B'}$. In this way, statement~(\textit{i}) can be equivalently reformulated as: for any choice of self-adjoint operators $\{Y^i_{B'} \}$,
	\[
	\max_{\Psi:\text{ CPTP}}\sum_i\Tr{\Psi(\rho^i_B)\;Y^i_{B'}}\ge\sum_i\Tr{\sigma^i_{B'}\;Y^i_{B'}}\;.
	\]
	The final manipulation amounts to shift and rescale the operators $Y^i_{B'}$ so that they become all positive semi-definite and sum up to $I_{B'}$, namely,
	\[
	Y^i_{B'}\mapsto \frac{1}{\lambda}\left(Y^i_{B'}-\frac{1}{\mu}\sum_iY^i_{B'}+\nu I_{B'}\right)\;,
	\]
	for suitable constants $\lambda,\mu,\nu>0$. This can always be done without loss of generality, due to the fact that $\Tr{\Psi(\rho^i_B)}=\Tr{\rho^i_B}=\Tr{\sigma^i_{B'}}$, for all $i$.
	
	Summarizing, until now we have shown that statement~(\textit{i}) is equivalent to the following: for any POVM $\{P^i_{B'} \}$,
	\begin{equation}\label{eq:summ}
	\max_{\Psi:\text{ CPTP}}\sum_i\Tr{\Psi(\rho^i_B)\;P^i_{B'}}\ge\sum_i\Tr{\sigma^i_{B'}\;P^i_{B'}}\;.
	\end{equation}
	
	Let us know introduce a quantum system $\bar{B}$ such that $\sH_{\bar{B}}\cong\sH_{B'}$.
	Recalling the property of the maximally entangled state, for which
	\[
	\Tr{x_{B'}y_{B'}}=d_{B'}\<\Phi^+_{\bar{B}B'}|(x_{\bar{B}}\otimes y_{B'}^T)|\Phi^+_{\bar{B}B'}\>\;,
	\]
	where the transposition is meant with respect to the basis fixed when writing $|\Phi^+_{\bar{B}B'}\>$, we can rewrite condition~(\ref{eq:summ}) as:
	\begin{align*}
	&\max_{\Psi:\text{ CPTP}}\<\Phi^+_{\bar{B}B'}|(\Gamma^\dagger_{\bar{A}}\otimes\Psi_B)(\rho_{\A B})|\Phi^+_{\bar{B}B'}\>\\
	&\ge\<\Phi^+_{\bar{B}B'}|(\Gamma^\dagger_{\bar{A}}\otimes\id_{B'})(\sigma_{\A B'})|\Phi^+_{\bar{B}B'}\>\;,
	\end{align*}
	where $\Gamma^\dagger:\bar{A}\to \bar{B}$ is a unital\footnote{That is, $\Gamma^\dagger(I_\A)=I_{\bar{B}}$.} CP map defined by the relation
	\[
	\Gamma^\dagger(Z_\A)=\sum_i\Tr{Z_\A\;\omega^i_{\A}}P^i_{\bar{B}}\;,
	\]
	where $\omega^i_{\A}$ are the states we fixed at the beginning of the proof and $\{P^i_{\bar{B}}\}_i$ is an arbitrary POVM. Using once more the ``ricochet property'' of the maximally entangled state, and recalling the fact that $\rho_{\A B}$ and $\sigma_{\A B'}$ are Choi operators, that is,
	\begin{align*}
		&(\Gamma^\dagger_{\bar{A}}\otimes\Psi_B)(\rho_{\A B})\\
		&=(\Gamma^\dagger_{\bar{A}}\otimes\Psi_B\circ\mN)(\Phi^+_{\A A})\\
		&=(\id_{\bar{B}} \otimes \Psi_B \circ\mN\circ \overline{\Gamma}_{B'} )(\Phi^+_{\bar{B} B'})\;,
	\end{align*}
	where $\overline{\Gamma}_{B'}$ denotes the complex conjugation of $\Gamma_{B'}$ (namely, a CPTP map), we finally obtain the following reformulation of statement~(\textit{i}): for any measure-and-prepare CPTP map $\Gamma:B'\to A$ (recall that $\bar{B}\cong B'$ and $\bar{A}\cong A$),
	\begin{align*}
	&\max_{\Psi:\text{ CPTP}}\<\Phi^+_{\bar{B}B'}|(\id_{\bar{B}}\otimes\Psi_B\circ\mN_A\circ\Gamma_{B'})(\Phi^+_{\bar{B}B'})|\Phi^+_{\bar{B}B'}\>\\
	&\ge\<\Phi^+_{\bar{B}B'}|(\id_{\bar{B}}\otimes\mN'_A\circ\Gamma_{B'})(\Phi^+_{\bar{B}B'})|\Phi^+_{\bar{B}B'}\>\;.
	\end{align*}
	
	A \textit{sufficient} condition for statement~(\textit{i}) is hence given by the following:
	for any measure-and-prepare CPTP map $\Gamma:B'\to A$,
	\begin{align*}
	&\max_{\Psi:\text{ CPTP}}\<\Phi^+_{\bar{B}B'}|(\id_{\bar{B}}\otimes\Psi_B\circ\mN_A\circ\Gamma_{B'})(\Phi^+_{\bar{B}B'})|\Phi^+_{\bar{B}B'}\>\\
	&\ge\max_{\Psi':\text{ CPTP}}\<\Phi^+_{\bar{B}B'}|(\id_{\bar{B}}\otimes\Psi'_{B'}\circ\mN'_A\circ\Gamma_{B'})(\Phi^+_{\bar{B}B'})|\Phi^+_{\bar{B}B'}\>\;.
	\end{align*}
	The above equation, using the results of Ref.~\cite{KRS09}, becomes
	\[
	\hmin(\bar{B}|B)_{(\id\otimes\mN\circ\Gamma)\Phi^+}\le	\hmin(\bar{B}|B')_{(\id\otimes\mN'\circ\Gamma)\Phi^+}\;,
	\]
	for all measure-and-prepare encoding CPTP maps $\Gamma$.
	
	On the other hand, at this point it is easy to see that the condition above is not only necessary, but also sufficient for statement~(\textit{i}) to hold, due to the data-processing theorem satisfied by conditional min-entropy (see, e.g., Ref.~\cite{Toma-book}).
	
\medskip\section{Conclusions}

We showed how the three orderings of degradability, ambiguity, and coherence are equivalent, whereas the orderings of degradability and noisiness are not. The proof has been presented in full generality, encompassing classical channels, quantum channels with classical-quantum encodings, and quantum channels with fully general quantum encodings. In~\cite{shannon_1958}, Shannon considers another ordering between channels, namely, ``channel inclusion.'' This is similar to the degradability condition, in the sense that it is based on the notion of simulability (of one channel by means of another), however, the set of transformations allowed is much more general, as it includes general encodings, decodings, and free use of shared randomness. Channel inclusion too can be studied from the general viewpoint of statistical comparison: this has been done in the classical scenario in~\cite{nasser1,nasser2}, whereas, in the quantum scenario, a similar ordering, in which not only shared randomness but also forward classical communication is freely available, has been studied in~\cite{RBL17}.

\section*{Acknowledgment}

This work was supported in part by JSPS KAKENHI, grants no. 26247016 and no. 17K17796.

% trigger a \newpage just before the given reference
% number - used to balance the columns on the last page
% adjust value as needed - may need to be readjusted if
% the document is modified later
%\IEEEtriggeratref{8}
% The "triggered" command can be changed if desired:
%\IEEEtriggercmd{\enlargethispage{-5in}}

% references section

% can use a bibliography generated by BibTeX as a .bbl file
% BibTeX documentation can be easily obtained at:
% http://mirror.ctan.org/biblio/bibtex/contrib/doc/
% The IEEEtran BibTeX style support page is at:
% http://www.michaelshell.org/tex/ieeetran/bibtex/
%\bibliographystyle{IEEEtran}
% argument is your BibTeX string definitions and bibliography database(s)
%\bibliography{IEEEabrv,../bib/paper}
%
% <OR> manually copy in the resultant .bbl file
% set second argument of \begin to the number of references
% (used to reserve space for the reference number labels box)

\IEEEtriggeratref{3}

% that's all folks
\end{document}